\documentclass[iop,english,twocolappendix,numberedappendix,showpacs,superscriptaddress,appendixfloats,tighten,apj, twocolumn]{aastex6}
\usepackage{amsmath,amssymb,color,latexsym,natbib,graphicx}
\usepackage{dcolumn}
\usepackage{bm}
\def\ADD#1{{\textcolor{black}{#1}}}    

\usepackage{amstext}
\usepackage{graphicx}
\usepackage{babel}
\usepackage{hyperref}
\usepackage{color}
\usepackage{ulem}
\def\ADDD#1{{\textcolor{black}{#1}}}    

\def\bb{{\bf b}}
\def\vv{{\bf v}}
\def\vA{{\bf v_A}}

\def\xx{{\bf x}}
\def\zz{{\bf z^\pm}}
\def\zzp{{\bf z^+}}
\def\zzm{{\bf z^-}}
\def\be{\begin{equation}}
\def\ee{\end{equation}}
\def\ba{\begin{eqnarray}}
\def\ea{\end{eqnarray}}

\def \pmbtext#1{\leavevmode
     \setbox0\hbox{#1}
     \kern0,4pt \copy0 \kern-\wd0
     \kern-0,2pt \raise0,3pt \box0 }

\begin{document}
\title{Scaling of compressible magnetohydrodynamic turbulence in the fast solar wind}
\author{S. Banerjee\altaffilmark{1,2}, L. Z. Hadid\altaffilmark{2}, F. Sahraoui\altaffilmark{2}, and S. Galtier\altaffilmark{3}}
\email{biswayan@gmail.com}
\affil{$^1$ Universit\"at zu K\"oln, Institut fur Geophysik und Meteorologie, Pohligstrasse 3, 50969 K\"oln, Germany}
\affil{$^2$ LPP, CNRS-Ecole Polytechnique-UPMC-Universit\'e Paris-Sud, F-91128 Palaiseau, France}
\affil{$^3$ LPP, Universit\'e Paris-Sud, Ecole Polytechnique, F-91128 Palaiseau, France}

\begin{abstract}   
The role of compressible fluctuations in the energy cascade of fast solar wind turbulence is studied using a reduced form of
an exact law derived recently \citep{Banerjee13} for compressible 
isothermal magnetohydrodynamics and {\it in-situ} observations from the THEMIS B/ARTEMIS P1 spacecraft. A statistical survey of the data revealed a turbulent energy cascade over two decades of scales, which is broader than the previous estimates made from an exact incompressible law. 
A term-by-term analysis of the compressible model reveals new insight into the role played by the compressible fluctuations in the energy cascade.
The compressible fluctuations are shown to amplify (2 to 4 times) the turbulent cascade rate with respect to the incompressible model in $\sim 10\%$ of the analyzed samples. 
This new estimated cascade rate is shown to provide the adequate energy dissipation required to account for the local heating of the non-adiabatic solar wind. 
\end{abstract}
\maketitle

\section{Introduction}\label{sec:Intro}
The solar wind is an excellent natural laboratory for the {\it in situ} study of space plasma turbulence~\citep{Bruno05,jltp}. Due to the relatively weak density 
fluctuations ($\sim 10\%$), fast solar wind (FSW) turbulence is often described at low frequencies ($< 0.1$\,Hz) by incompressible magnetohydrodynamics 
(MHD)~\citep{Goldstein,luca,galtier12}. 
\ADDD{However, the high correlation between the velocity and the magnetic field in the FSW leads to a strong imbalance between the outward and inward propagating 
Alfv\'en waves, which in turn makes the incompressible nonlinear cascade small. A compressible cascade may overcome this problem and explain
the turbulent character of the FSW. Furthermore, it may provide a natural source for a local heating which is required in order to understand the slow decrease 
of the solar wind temperature with the heliospheric distance~\citep{Marsch,Vasquez07}.}
The pioneering works~\citep{BB89b, MT90} included attempts to understand the origin and the nature of the density fluctuations, as well as their spectral laws. 
A Kolmogorov--like $-5/3$ spectrum for the density fluctuations led to the conclusion that the density acts as a passive scalar in the solar wind. \ADD{However, in the following years, several studies
explored the plausibility of an active participation of the density fluctuations using parametric decay of solar wind turbulence \citep{Grappin90, Malara00}.
More recently, a study by \cite{Hnat05} showed that the scaling of extended self-similarity of the density fluctuations does not coincide with that expected 
for a passive scalar (e.g., the magnetic field magnitude for incompressible MHD turbulence).}

A direct evidence of the presence of an inertial energy cascade in the solar wind was observed using the so-called Yaglom law \citep{luca,Marino08}. It is a 
universal law derived analytically from the incompressible MHD equations \citep{PP98a} (hereafter PP98) under the assumptions of homogeneity, stationarity 
and isotropy of the turbulent fluctuations. 
Later, a first attempt was made to include the compressibility using a heuristic model \citep{Carbone09,Marino11} (hereafter C09). The application of C09 to 
FSW turbulence showed a better scaling relation of the energy flux than with PP98. Furthermore, a significant increase of the turbulent cascade rate was 
evidenced and was shown to be sufficient to account for the local heating of the non-adiabatic solar wind expansion \citep{Carbone09}. 
\ADDD{Although those results are original and constitute a real leap forward in studies of solar wind turbulence, (i) C09 remains a heuristic model 
(i.e., not derived analytically as the exact law of PP98), and gives a different origin of the amplification of the energy cascade rate than the one evidenced in the present work,
(ii) following incompressible MHD turbulence, C09 attempted to verify two scaling relations corresponding to two pseudo-energy conservations, however, in compressible turbulence only the total energy is conserved (not the individual pseudo-energies) \citep{Marsch87, Banerjee13}, and (iii) the frequency range chosen for the study does not seem to correspond fully to the MHD inertial range \citep{Forman10, Lucasreply}.}

In this Letter, we present for the first time a statistical study of scaling properties of FSW turbulence \ADDD{using 
a reduced form of an exact law} derived recently by \cite{Banerjee13} (hereafter BG13) for compressible isothermal MHD turbulence (see also \cite{GB11}). 
Our findings show the new role played by the compressible fluctuations in the turbulent cascade and the local heating of the FSW.

\section{DIFFERENT MODELS} \label{sec:Results} 
In the course of this Letter, we shall compare two solar wind turbulent MHD models, namely the incompressible PP98 and the compressible isothermal BG13 exact laws. For the sake of clarity, we recall their different relationships written for the dissipation rate of the total energy.
\ADDD{We recall that these laws are derived under the assumptions of a homogeneous, stationary turbulence, and in the asymptotic limit of large kinetic and magnetic Reynolds numbers.}

\paragraph*{\textbf{Incompressible model}:}  The PP98 law is written in terms of the Els\"asser variables $\zz = \vv \pm \vA$, where $\vv$ is the flow velocity, $\vA \equiv \bb/\sqrt{\mu_0 \rho}$ is the magnetic field normalized to a velocity and $\rho$ is the plasma density (in this incompressible model, we take $\rho=\langle \rho \rangle$). It reads (in the isotropic case)
\be
- \frac{4}{3} \varepsilon_{I} \ell = 
\left\langle {\left(\delta \zzp \right)^2 \over 2} \delta z_{\ell}^- + {\left(\delta \zzm \right)^2 \over 2} \delta z_{\ell}^+ \right\rangle \langle \rho \rangle 
\equiv {\cal F}_I (\ell) \, , \label{pp98a}
\ee
where the general definition of an increment of a variable $\psi$ is used, i.e. $\delta \psi \equiv \psi (\xx + \boldsymbol{\ell}) - \psi (\xx)$. 
The longitudinal components are denoted by the index $\ell$ with $\ell \equiv \vert \boldsymbol{\ell} \vert $, $\langle \cdot \rangle$ stands for the statistical average and $\varepsilon_{I}$ is the dissipation rate of the total energy. Note that in S.I. units, we have the relation $\left\langle \rho\right\rangle = 1.673 \times 10^{-21}  \left\langle n_p \right\rangle $. 

\paragraph*{\textbf{Compressible model}:} The exact law BG13 can schematically be written as
\be
- 4 \varepsilon_C = \nabla_{\boldsymbol{\ell}}  \cdot {\cal F}_C -  \Phi + {\cal S} \, ,
\label{bg13relation}
\ee
where $\varepsilon_C$ denotes the dissipation rate of the total compressible energy. The flux term writes
\ba
{\cal F}_C &= \left\langle \frac{1}{2} \left[  \delta ( \rho  \mathbf{z}^+) \cdot \delta \mathbf{z}^+ \right]  {\delta \mathbf{z}^-} 
+\frac{1}{2} \left[ \delta ( \rho \mathbf{z}^-) \cdot \delta \mathbf{z}^-  \right]  {\delta  \mathbf{z}^+}\right\rangle \label{equ4} \\
&+ \left\langle 2  \delta \rho \delta e  \delta \vv \right\rangle 
+ \left\langle 2 {\overline{\delta} \left(e + { v_A^2  \over 2}\right) \delta ( \rho_1 \vv)}  \right\rangle \, , \nonumber
\ea
where by definition $\overline{\delta} \psi \equiv (\psi (\xx + \boldsymbol{\ell}) + \psi (\xx))/2$ and $e$ is the internal energy 
($e=c_s^2 \ln (\rho / \langle \rho \rangle)$, with $c_s$ the constant isothermal sound speed, $\langle \rho \rangle$ the mean density and $\rho = \langle \rho \rangle + \rho_1$). 
Note that ${\cal F}_C$ reduces to ${\cal F}_I$ when $\rho_1 =0$ (implying also that $\delta \rho =0$).
Furthermore, we have 
\be
\label{equ5}
\Phi =  \left \langle { 1 \over \beta' } \nabla' \cdot ( \rho \mathbf{v} e')  +  {1 \over \beta} \nabla \cdot  ( \rho' \mathbf{v}' e) \right \rangle \, ,
\ee
where the primed and unprimed variables correspond to the variables at points $\xx + \boldsymbol{\ell}$ and $\xx$ respectively and $\beta = 2 c_s^2 / v_A^2$ 
gives the local ratio of thermal to magnetic pressure (note the difference between the definition of $\beta$ used here and in \cite{Banerjee13}).
The last term ${\cal S}$ is a source term that includes the local divergences of $\vv$ and $\vA$.
The main goal of this study is to evaluate for the first time the compressible effects in solar wind turbulence with an exact law. This objective will 
be partly achieved by evaluating the first two terms in the right hand side of Equation (\ref{bg13relation}). The source term will be left aside because a 
reliable evaluation of local velocity divergences is not possible using single spacecraft data. Thus, we implicitly assume that ${\cal S}$ is subdominant. 
Note that this situation, not proved for the solar wind, is well observed numerically in supersonic HD turbulence \citep{kritsuk13} and in a preliminary study using numerical simulations of isothermal MHD turblence \citep{Servidio15}.

We may try to estimate $\Phi$, which is not a pure flux term but can be reduced to it, if the \ADDD{plasma $\beta$ is relatively stationary}. In this particular case, we obtain after simple manipulations
\be
\Phi = {1 \over \beta} \nabla_{\boldsymbol{\ell}} \cdot  \left\langle \rho \mathbf{v} e' -  \rho' \mathbf{v}' e \right\rangle 
= - {2 \over \beta} \nabla_{\boldsymbol{\ell}} \cdot \left\langle  {\overline{\delta} e \delta ( \rho \mathbf{v})} \right\rangle \, . 
\ee
This term can now be merged with the flux terms in Equation~(\ref{equ4}). This results in modifying the last term of ${\cal F}_C$ from 
${\overline{\delta} (e + { v_A^2  / 2}) \delta (\rho_1 \vv)}$ to ${\overline{\delta} [(1 + \beta^{-1})e + { v_A^2 / 2}] \delta (\rho_1 \vv)}$. 
As a last step, one can integrate relation~(\ref{bg13relation}) over a ball of radius $\ell$ and get the equivalent of the isotropic relation (\ref{pp98a}) for isothermal compressible MHD turbulence, namely 
\be
-{4 \over 3} \varepsilon_C \ell = {\cal F}_{C+\Phi}(\ell) \, , \label{bg13reduced}
\ee
where 
\be
{\cal F}_{C+\Phi}(\ell) = {\cal F}_{1}(\ell) + {\cal F}_{2}(\ell)+{\cal F}_{3}(\ell) \, , 
\ee
and
\ba
{\cal F}_{1}(\ell) &=&\left\langle \frac{1}{2} \left[ \delta ( \rho \mathbf{z}^-) \cdot \delta \mathbf{z}^-  \right]  {\delta  {z}_{\ell}^+}
+  \frac{1}{2} \left[  \delta ( \rho  \mathbf{z}^+) \cdot \delta \mathbf{z}^+ \right]  {\delta {z}_{\ell}^-} \right\rangle \, , \nonumber \\
{\cal F}_{2}(\ell)&=& \left\langle 2  \delta \rho \delta e  \delta v_{\ell} \right\rangle \, , \nonumber \\
{\cal F}_{3}(\ell)&=& \left\langle 2 {\overline{\delta} \left[ \left(1 + \frac{1}{\beta} \right) e + { v_A^2  \over 2}\right] \delta ( \rho_1 v_{\ell})}  \right\rangle \, . 
\label{fcphi}
\ea
Equations~(\ref{bg13reduced})--(\ref{fcphi}) will be evaluated using spacecraft data in the FSW. It is worth noting that the condition of \ADDD{uniform} 
$\beta$ used to obtain the new form of ${\cal F}_{3}(\ell)$ in Equations~(\ref{fcphi}) is a stringent requirement in selecting the data used in the present study. \ADDD{We note however that it is the local $\beta$ that is used in evaluting the flux terms and not its mean.}

\section{Estimation of the energy cascade rates}\label{sec:Results} 
\subsection{Data selection} 
\begin{figure}
        \begin{centering}
                \includegraphics[width=1\columnwidth]{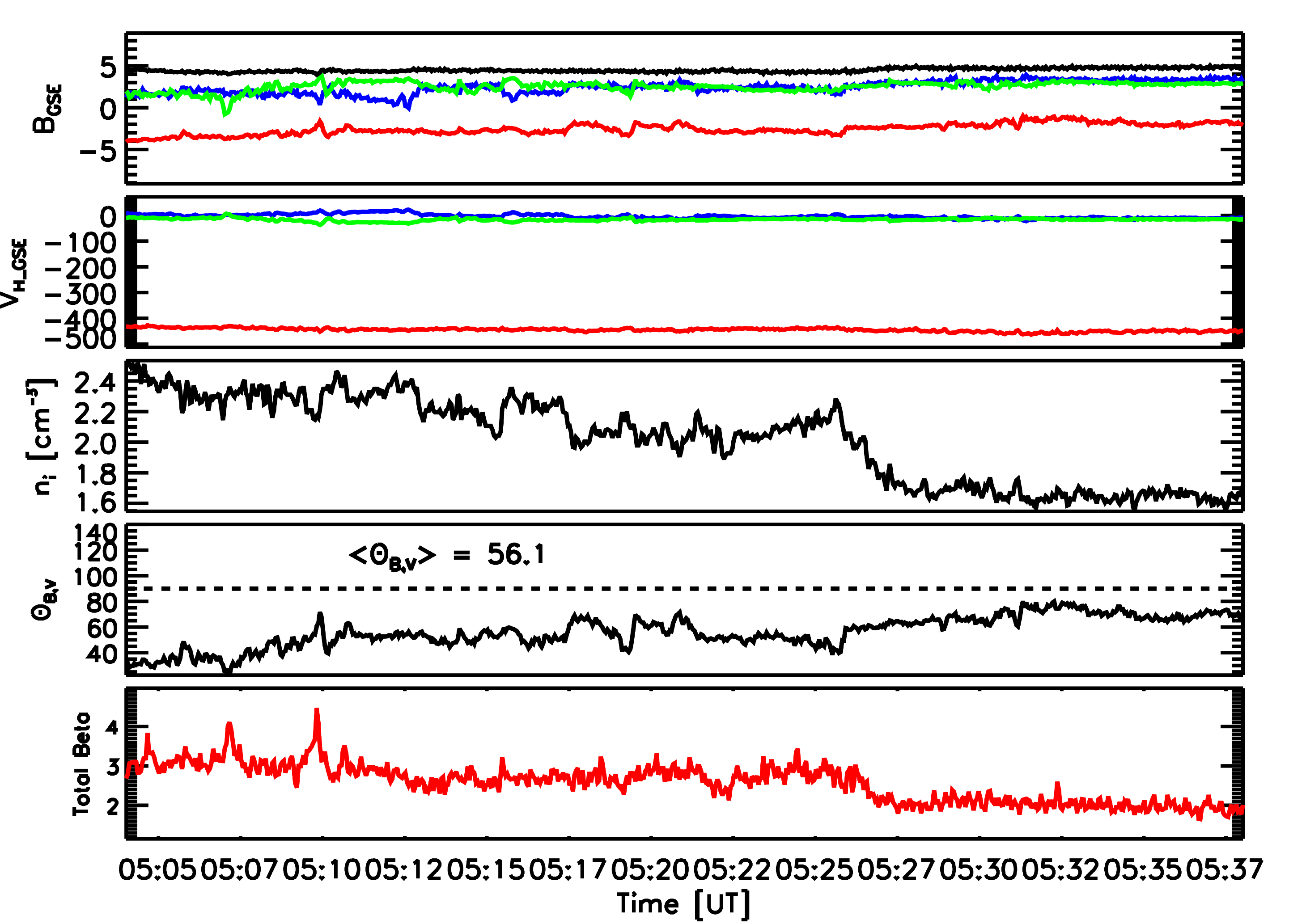}\par
        \end{centering}
        \caption{From top to bottom: the solar wind magnetic field components, ion velocity, ion number density, $\Theta_{\bf VB}$ angle and total plasma beta 
        ($\beta=\beta_i+\beta_e$) measured by the FGM and ESA experiments onboard the THEMIS B spacecraft on day 2008-06-29 from 05:02 to 05:37.
        }
\label{waveforms}  
\end{figure}

\begin{figure}
        \begin{centering}
                \includegraphics[width=1\columnwidth]{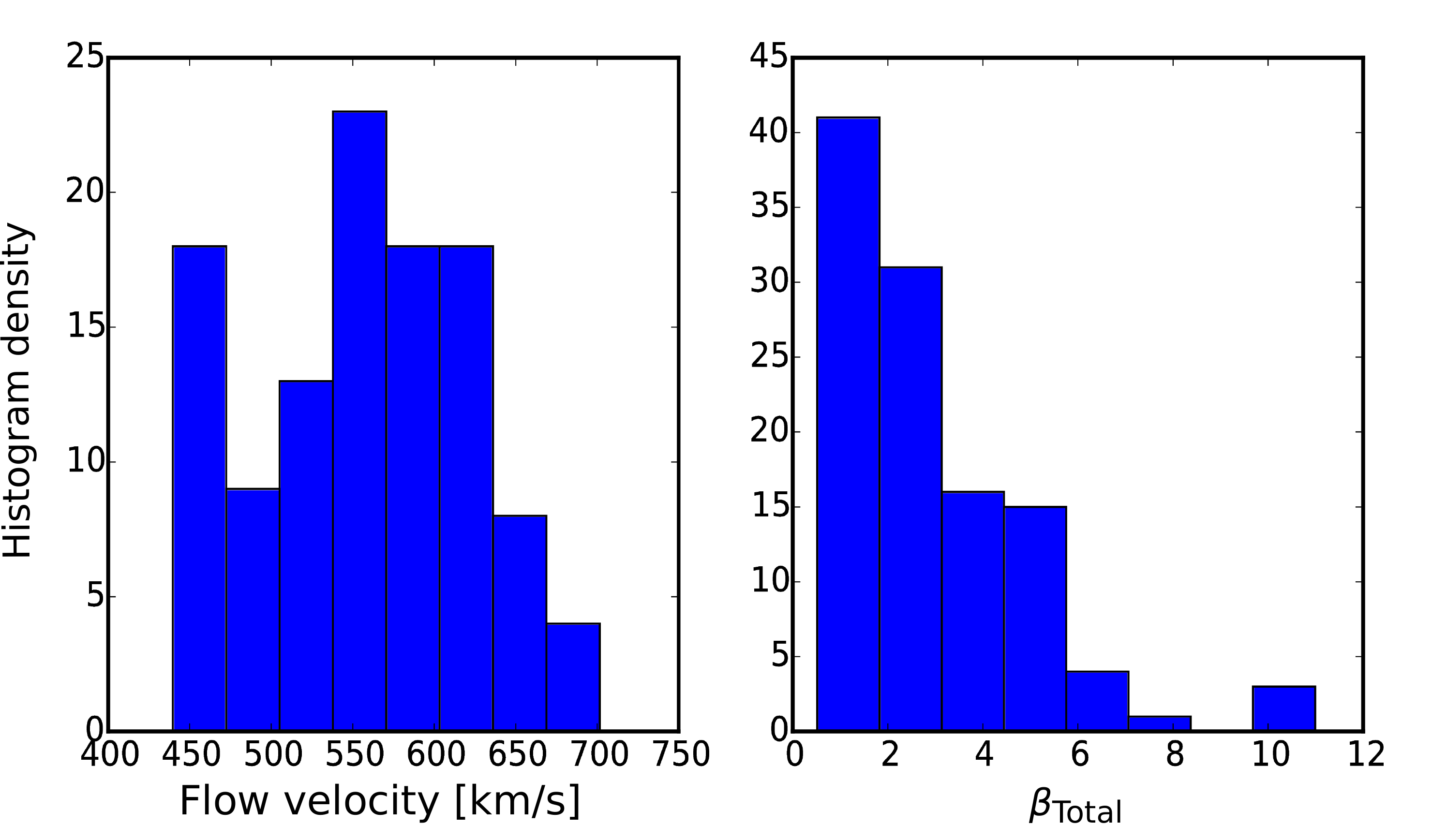}\par
        \end{centering}
       \caption{Histograms of the solar wind speed (left) and total plasma $\beta$ (right) for all the statistical samples.
       Each sample represents $35$mn ($\sim 700$  points) of data.}
\label{histogram_data}
\end{figure}

We used the THEMIS B/ARTEMIS P1 spacecraft data during time intervals when it was travelling in the free-streaming solar wind. In particular, we used the plasma moments and magnetic field data
which were measured respectively by the Electrostatic Analyzer (ESA) and the Flux Gate Magnetometer (FGM) with a time resolution of 3 seconds (i.e., spin period). Since we are interested in FSW, \ADDD{we selected a total of 148 
intervals} between 2008 and 2011 for which $ V_{sw} > 450 \ \text{km} \, \text{s}^{-1}$,  where $V_{sw}$ is the solar wind speed. Furthermore, we have tried as much as possible to avoid data intervals that contained significant 
ecliptic disturbances, such as coronal mass ejection or interplanetary shocks. Besides these criteria, we paid a particular attention to choosing only intervals that showed relatively stationary plasma $\beta$ and $\Theta_{\bf VB}$,
the angle between the local solar wind speed ${\bf V}$ and the magnetic field ${\bf B}$. The stationarity of the plasma $\beta$ is imposed to fulfill the condition used to derive Equations~(\ref{bg13reduced})--(\ref{fcphi}), as discussed
in the previous section. The stationarity of the angle $\Theta_{\bf VB}$ is required to guarantee that the spacecraft is sampling nearly the same direction of space with respect to the local magnetic field (when the Taylor hypothesis, $\ell \sim V \tau$, is used), which would
ensure a better convergence in estimating the cascade rate. Indeed, if the angle $\Theta_{\bf VB}$ changes significantly in a single time interval (e.g., from $\sim 0^\circ$ to $\sim 90^\circ$), this means that the analysis would mix
between the two cascade rates estimated along the direction parallel and perpendicular to the local magnetic field, known to be very different. This is based on anisotropic MHD turbulence models and on spacecraft observations in the solar 
wind~\citep{macbride08} (this point will be discussed in more detail in an upcoming paper). The obtained intervals that fulfilled all the previous criteria were divided into a series of samples of equal duration $\sim 35$mn, which
corresponds to $\sim 700$ data points with a $3$s time resolution. This sample size is much larger than those used in previous studies based on ACE spacecract data that had a time resolution of $24$s (e.g., ~\cite{macbride08}). The sample
size of $35$mn ensures having at least one correlation time of the turbulent fluctuations estimated to vary in the range $\sim 20-30$mn. The data selection yielded 170 samples and a total number of data points $\sim 14\times10^4$. An
example of the analyzed time intervals is shown in Figure~\ref{waveforms}.  The average solar wind speed and plasma $\beta$ for all the statistical samples are shown in Figure~\ref{histogram_data}. 

\subsection{Results} 
We have constructed temporal structure functions of the different turbulent fields involved in the BG13 exact law at different time lags $\tau$ and verified their linear scaling with respect to $\tau$. In order to probe into the scales of the inertial range, known to lie within the frequency range $\sim$[$10^{-4},1$]\,Hz (based on the observation of the Kolmogorov-like $-5/3$ magnetic energy spectrum, \cite{Bruno05,Marino08}), we vary the time lag $\tau$ from 10\,s to 1000\,s thereby being well inside the targeted frequency range.

\begin{figure}
        \begin{centering}
                \includegraphics[width=1\columnwidth]{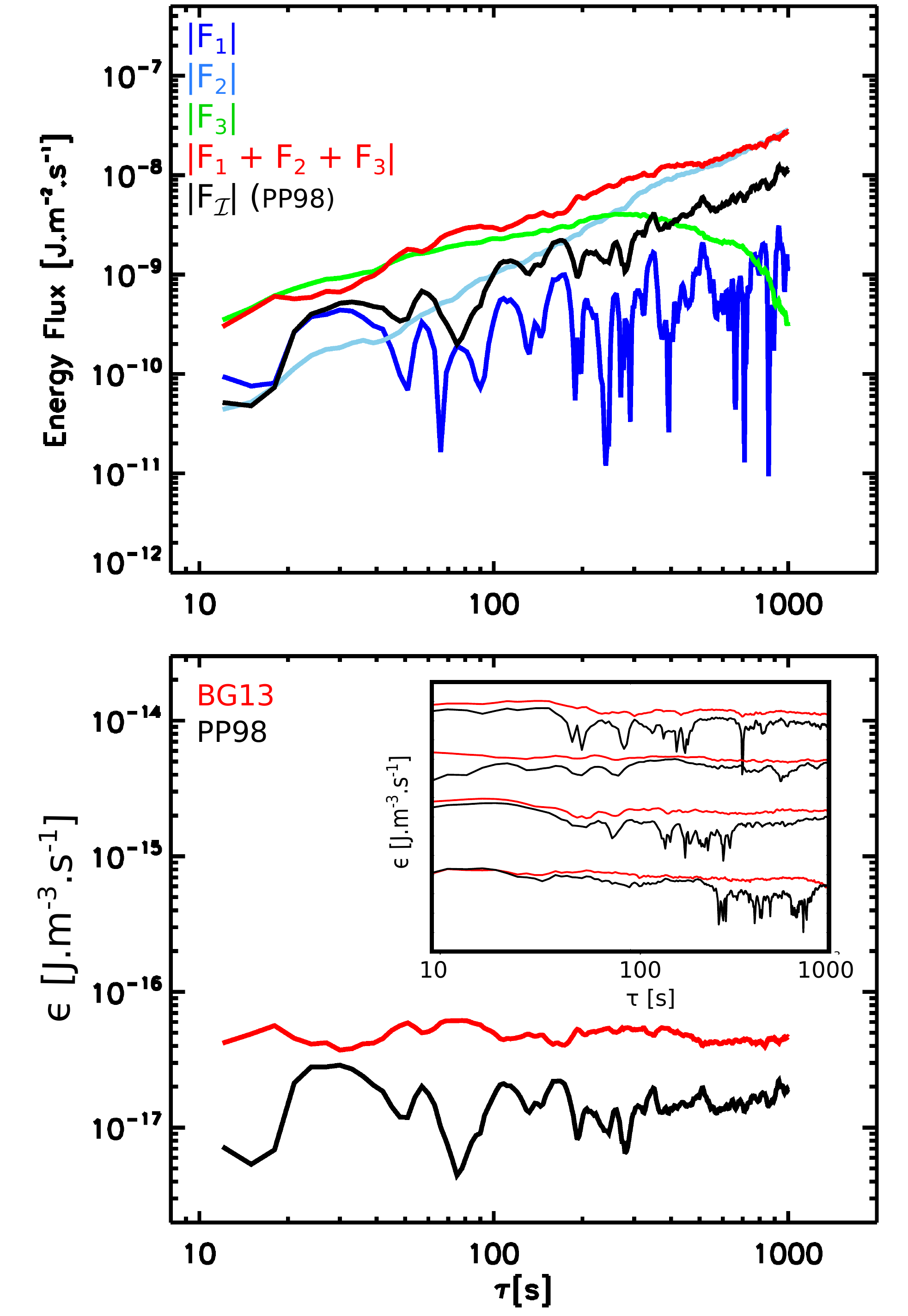}\par
        \end{centering}
       \caption{Top: comparison of the different terms $\vert {\cal F}_1 \vert$, $\vert {\cal F}_2 \vert$ and $\vert {\cal F}_3 \vert$ (see text for the definitions) of the flux 
       ${\cal F}_{C+\Phi}$. Bottom: comparison between the turbulent cascade rates given by the PP98 and BG13 models. The compressibility is $13.7\%$.
       The inset shows other examples for which BG13 model gives a smoother cascade rate over two decades than the PP98 model.}
\label{flux_terms_examp}
\end{figure}

Figure~\ref{flux_terms_examp} (bottom) shows for a few case study (data from Fig.~\ref{waveforms}) a comparison between the energy cascade rates $\varepsilon_{I,C}$ of the incompressible and the compressible models. They were estimated using expressions (\ref{pp98a}) and (\ref{bg13reduced}). The compressibility, defined as $\sqrt{\left( \langle \rho^2  \rangle - \langle \rho \rangle^2  \right)}/ \langle \rho \rangle$, is about $14\%$. One can see that the energy cascade rate from BG13 gives a smoother scaling than the PP98 model over two decades of (time) scales $\tau$, which defines in a more rigorous way the size of the inertial range. This behaviour is representative of most of the other studied intervals, as can be seen from the few cases shown in the inset of Figure~\ref{flux_terms_examp} (bottom). 
The value corresponding to the plateau gives an estimate of the rate of the total energy dissipation per unit volume \citep{Vasquez07,Marino08}. In the case of the isothermal compressible law, we obtain $\vert \varepsilon_C \vert  \sim 6\times 10^{-17}$J m$^{-3}$s$^{-1}$. The estimate from the incompressible law gives a value about 
$3$ times smaller. 

In order to quantify the contribution of the different compressible fluctuations, we show in Figure~\ref{flux_terms_examp} (top) the different flux terms ${\cal F}_1$, ${\cal F}_2$ and ${\cal F}_3$ separately. Note that the flux ${\cal F}_1$ can be seen as the generalization to the compressible case of the PP98 flux since it converges to it in the incompressible limit. For that reason we call it the Yaglom flux. We clearly see that the main contribution comes from the new pure compressible fluxes ${\cal F}_2$ and ${\cal F}_3$ with 
up to an order of magnitude of difference with ${\cal F}_1$.

\begin{figure}
        \begin{centering}
                \includegraphics[width=1\columnwidth]{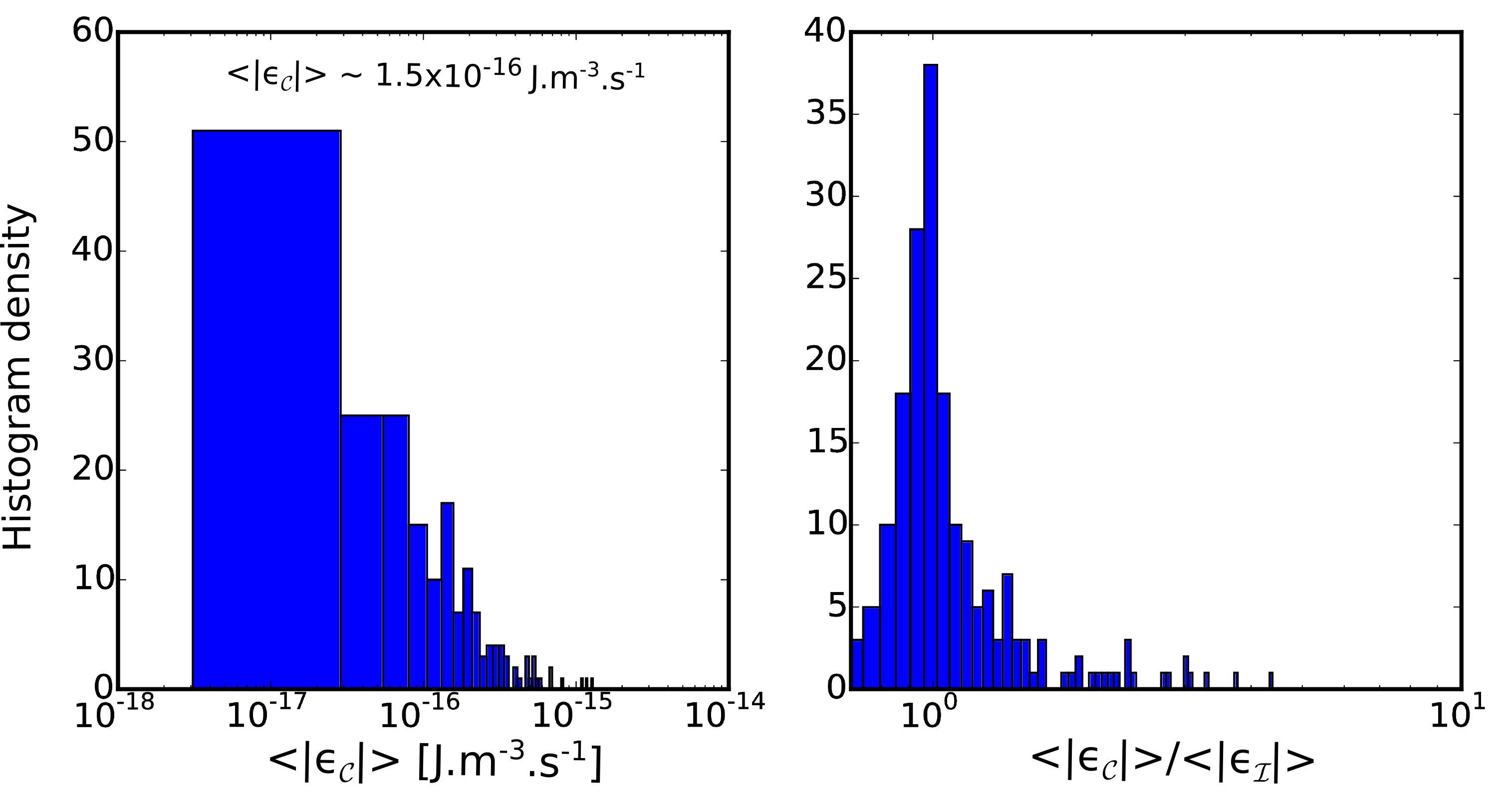}\par
        \end{centering}
       \caption{Left: histrogram of the average cascade rate for all the analyzed intervals. Right: histrogram of the ratio between the compressible (BG13) and incompressible (PP98) cascade rates.}
\label{histogram_epsilon}
\end{figure}
These results are confirmed in the statistical survey of all the samples. Figure~\ref{histogram_epsilon} (right) compares the ratio between the estimated cascades rates from the PP98 and the BG13 models. It is interesting to note that, while the compressible and incompressible models converge toward the same value of the cascade rate for most of the events, some cases show that the compressible rate is a few times larger than the incompressible one. These ratios remain however smaller than those reported in~\cite{Carbone09} (this point will be discussed elsewhere). 
The absolute values of the compressible cascade rate $\varepsilon_{C}$ (Fig.~\ref{histogram_epsilon} -- left) shows some spread around the mean value 
$\sim 1.5\times10^{-16}$J\,m$^{-3}$\,s$^{-1}$.

\begin{figure}
        \begin{centering}
               \includegraphics[width=1\columnwidth]{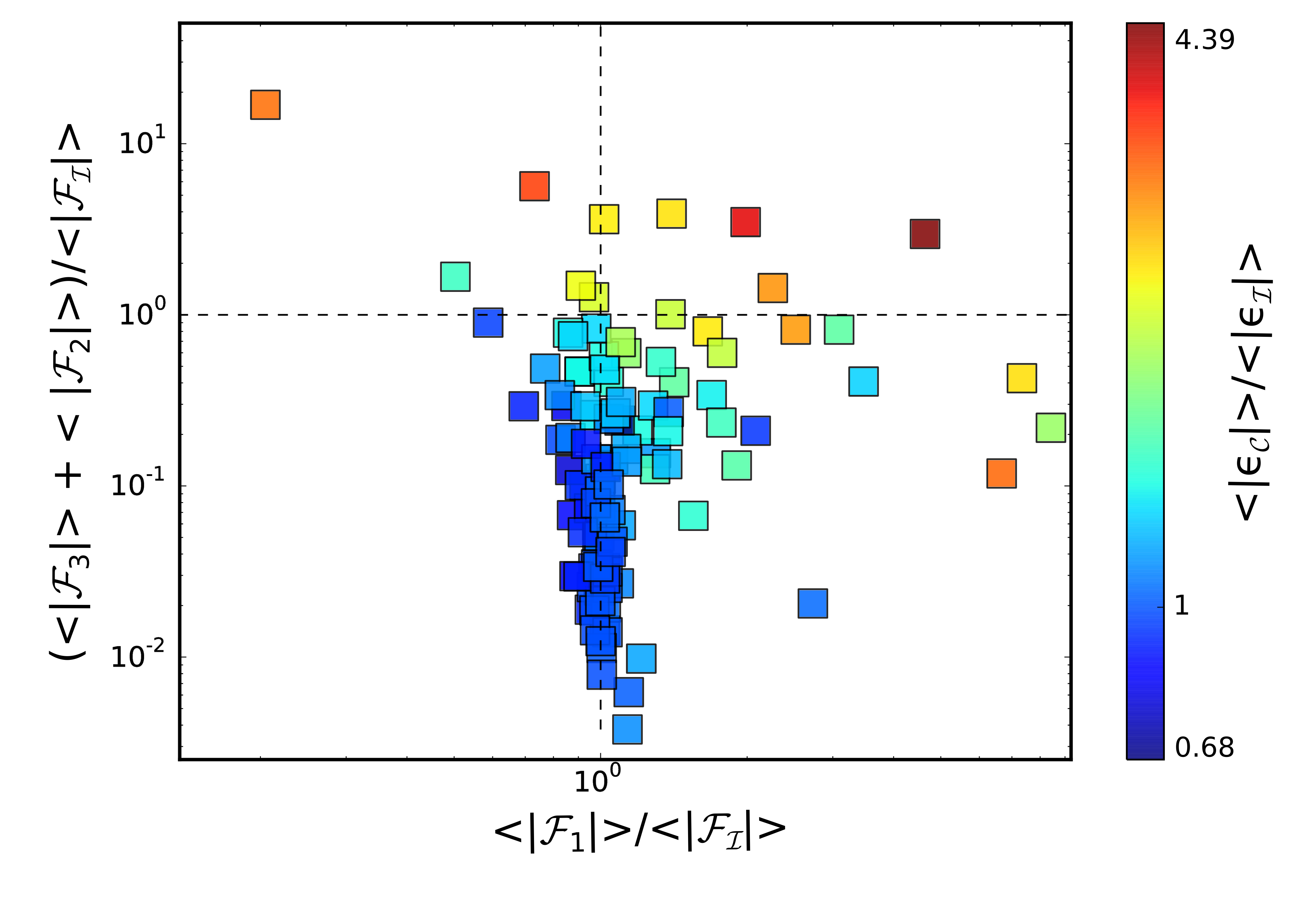}\par
        \end{centering}
       \caption{Estimation of the contribution of the compressible fluxes w.r.t. incompressible (Yaglom) flux to the total cascade rate.}
\label{epsilon_2D}
\end{figure}

More insight is gained when analyzing statistically the contribution of the different compressible fluxes, ${\cal F}_1$, ${\cal F}_2$ and ${\cal F}_3$, relative to the incompressible (Yaglom) flux ${\cal F}_I$. The result is shown in Figure~\ref{epsilon_2D}. A first observation is that most of the samples have their compressible Yaglom  
flux (${\cal F}_1$) of the order of the incompressible flux (${\cal F}_I$). This confirms the previous result that the density fluctuations entering into the compressible Yaglom flux does not play a leading role in amplifying the compressible cascade rate w.r.t. the incompressible one. This role is rather played by the new compressible fluxes ${\cal F}_2$ and ${\cal F}_3$: high values of $\langle \vert \varepsilon_{C} \vert \rangle/ \langle \vert \varepsilon_{I} \vert \rangle$ (up to $\sim 4$) are observed when 
$(\langle |{\cal F}_2| \rangle + \langle |{\cal F}_3| \rangle)/\langle \vert {\cal F}_I \vert \rangle>1$. Although a similar amplification has been reported in~\cite{Carbone09}, given by an heuristic modification of the incompressible (Yaglom) term {\it via} density fluctuations, the one observed here has a totally different origin: it is essentially due to the pure compressible terms ${\cal F}_2$ and ${\cal F}_3$ derived in the exact model of BG13. 
Note finally that the highest ratio $\langle \vert \varepsilon_{C} \vert \rangle / \langle \vert \varepsilon_{I} \vert \rangle$ (i.e., highest amplification of the cascade rate due to compressible fluctuations) is observed in the top-right quarter of Figure~\ref{epsilon_2D}, which corresponds to the cases when all the three terms ${\cal F}_1$, ${\cal F}_2$ and ${\cal F}_3$ dominate over the incompressible (Yaglom) term ${\cal F}_I$.

\section{DISCUSSION AND CONCLUSION} \label{conclusions} 
Unlike the incompressible PP98 model, the compressible flux obtained from BG13 model gives a uniform value of $\varepsilon_C$ estimated above over two decades of scales, thereby assuring a physical cascade process in fast solar wind turbulence. Using the heuristic model, \citet{Carbone09} found intervals for which either the inward or the outward flux scales linearly with the fluctuation scale. This problem has been overcome in the current study by using the flux of total energy which is an inviscid invariant of compressible MHD turbulence unlike the inward/outward flux separately. 

The turbulent cascade implies a forward flux of energy which ultimately will be converted at small-scales into heating by some kinetic processes~\citep[see e.g.][]{Sahraoui09,Sahraoui10}.
Using a simple power law model \citep{Vasquez07,Marino08}, we may obtain an estimate for the energy needed to heat up the fast solar wind at 1 AU. For a power-law of type $T (r) \sim r^{-\xi}$, with $T$ the proton temperature and $r$ the heliocentric distance, the model can be written as
\begin{equation}
\varepsilon_{h} = \frac{3}{2} \left( \frac{4}{3} - \xi \right) \frac{V_{sw} k_B T(r)}{m_p  r } \, , 
\label{heatm}
\end{equation}
where $\varepsilon_{h} $ is the energy flux rate (per unit mass). Using the average flow velocity $V_{sw}$ and temperature T for all the statistical events, with the value $\xi=0.49$ (corresponding to the upper bound of the estimated temperature using Ulysses data \citep{Marino08}) we estimated $\epsilon_h= 2.7\pm1.9\times 10^4 J kg^{-1}s^{-1}$. This value is of the order of the estimated energy cascade rate from BG13 model, $\left\langle \vert \epsilon_C \vert \right\rangle = 1.5\times10^{-16} J m^{-3}s^{-1} \sim 5.5\times 10^4 J kg^{-1}s^{-1}$ (using un average density $n_i = 1.7 \times 10^6 m^{-3}$), and also is in agreement with the finding of \citet{Carbone09}.  However, unlike the current study, a considerably low incompressible flux ($\equiv 10$ times smaller than the compressible flux) is reported in \citet{Carbone09} which can possibly be assigned to the absence of large scale drivers in the high latitude solar wind during solar minimum \citep{macbride08}.

However, this model can be improved in the future by using polytropic closure ~\citep{Banerjee14}, \ADD{taking the non-homogeneity and the expansion of the wind into account \citep{Verdini15} and also considering} the local anisotropy of the turbulence which is known to be 
important~\citep{Matthaeus90,Stawarz,narita,Sahraoui10,Osman,galtier12}. Previous studies~\citep{macbride08} have indeed shown that the heating is smaller in the parallel direction than in the perpendicular one, the latter being comparable however to the isotropic heating. A simple observational approach to account for anistropy would be to examine the dependence of the compressible cascade rate on the angle $\Theta_{\bf VB}$. 
A more complete approach would consist in splitting into two parts the flux term in Equation (\ref{bg13relation}) by assuming cylindrical isotropy around the local mean magnetic field direction. 
These problems will be investigated in a forthcoming paper where a detailed study of the nature of the cascade (direct {\it vs} inverse, inward {\it vs} outward) and a comparison between the fast and slow solar winds will be made (Hadid et al., in preparation). 

\vspace*{-0.3cm}
\paragraph*{Acknowledgment.}
The THEMIS/ARTEMIS data come from the AMDA data base (http://amda.cdpp.eu/).
We are grateful to Dr. O. Le Contel and Dr. L. Sorriso Valvo for useful discussions. FS acknowledges financial support from the ANR project 
THESOW, grant ANR-11-JS56-0008. The french participation in the THEMIS/ARTEMIS mission is funded by CNES and CNRS.

\end{document}